\documentclass{article} % For LaTeX2e
\usepackage[preprint]{colm2025_conference}

\usepackage{microtype}
\usepackage{hyperref}
\usepackage{graphicx}
\usepackage{url}
\usepackage{booktabs}
\usepackage{algorithm}
\usepackage{algpseudocode}
\usepackage{lineno}
\usepackage{multirow}
\usepackage{amsmath}

\definecolor{darkblue}{rgb}{0, 0, 0.5}
\hypersetup{colorlinks=true, citecolor=darkblue, linkcolor=darkblue, urlcolor=darkblue}

\title{Parallel Track Transformers: Enabling Fast GPU Inference with Reduced Synchronization}

\author{Chong Wang\thanks{Work done when authors were with Apple. Corresponding author: \nolinkurl{mr.chongwang@gmail.com}.}, Nan Du$^*$, Tom Gunter$^*$, Tao Lei, Kulin Seth, 
Senyu Tong$^*$, Jianyu Wang, \\
\textbf{Guoli Yin, Xiyou Zhou, Kelvin Zou$^*$, Ruoming Pang$^*$} \\
Apple}

\begin{document}

\ifcolmsubmission
\linenumbers
\fi

\maketitle

\begin{abstract}
Efficient large-scale inference of transformer-based large language models (LLMs) remains a fundamental systems challenge, frequently requiring multi-GPU parallelism to meet stringent latency and throughput targets. Conventional tensor parallelism decomposes matrix operations across devices but introduces substantial inter-GPU synchronization, leading to communication bottlenecks and degraded scalability. We propose the Parallel Track (PT) Transformer, a novel architectural paradigm that restructures computation to minimize cross-device dependencies. PT achieves up to a 16× reduction in synchronization operations relative to standard tensor parallelism, while maintaining competitive model quality in our experiments. We integrate PT into two widely adopted LLM serving stacks—TensorRT-LLM and vLLM—and report consistent improvements in serving efficiency, including up to 15-30\% reduced time to first token, 2-12\% reduced time per output token, and up to 31.90\% increased throughput in both settings.

% No throughput comparison version
% Efficient large-scale inference of transformer-based large language models (LLMs) remains a fundamental systems challenge, frequently requiring multi-GPU parallelism to meet stringent latency and performance requirements. Conventional tensor parallelism decomposes matrix operations across devices but introduces substantial inter-GPU synchronization, leading to communication bottlenecks and degraded scalability. We propose the Parallel Track (PT) Transformer, a novel architectural paradigm that restructures computation to minimize cross-device dependencies. PT achieves up to a 16× reduction in synchronization operations relative to standard tensor parallelism, with no degradation in model accuracy. We integrate PT into two widely adopted LLM serving stacks—TensorRT-LLM and vLLM—and report consistent improvements in serving efficiency, including reduced end-to-end latency and enhanced performance in both settings.

%We evaluate our approach in two serving environments—one with high-speed interconnects and another with slower interconnects—and demonstrate consistent performance improvements in both scenarios.
%\rp{Can we quantify the reduction of synchronization overhead and e2e performance improvements?}

%\cw{We can measure it on vllm or trt-llm}
\end{abstract}

\section{Introduction}

Transformer-based large language models (LLMs) have made rapid and significant progress in recent years~\citep{Vaswani+2017}. However, efficiently serving these models at scale remains a substantial challenge. Various parallelization strategies have been proposed to address this, including pipeline parallelism~\citep{huang2019gpipeefficienttraininggiant}, data parallelism~\citep{NIPS2012_6aca9700}, and tensor/model parallelism~\citep{shoeybi2020megatronlmtrainingmultibillionparameter}. Among these, tensor parallelism is particularly attractive for large models, as it enables distributed computation across multiple GPUs for the same inference request. Nevertheless, a key limitation of tensor parallelism lies in the frequent inter-GPU synchronizations required during inference—particularly for attention and feedforward layers—which can introduce significant latency and hinder scalability. As model sizes continue to increase, synchronization overhead has emerged as a critical bottleneck, constraining the efficiency and scalability of LLM serving infrastructure.

Several strategies have been proposed to reduce communication overhead in large-scale model training. Techniques such as overlapping communication with computation~\citep{chang2024fluxfastsoftwarebasedcommunication,zhang2025laddertransformer} aim to hide communication latency and improve overall resource utilization. Parallel transformer layers~\citep{gpt-j,chowdhery2022palmscalinglanguagemodeling} are designed to enable concurrent execution of attention and feedforward components, allowing the synchronization of their activations to occur simultaneously rather than sequentially. More recently,~\cite{kim2025spdsyncpointdropefficient} introduced a post-training method that selectively drops synchronization on attention outputs within tensor parallelism, achieving a favorable balance between reduced communication and minimal impact on model quality.

We introduce the Parallel Track (PT) Transformer, a novel model architecture designed to minimize inter-GPU synchronization overhead. Unlike traditional transformer models, PT consists of multiple transformer instances—referred to as "tracks"—that operate independently and in parallel across GPUs. Periodic synchronization is performed between tracks to preserve model quality while avoiding the frequent synchronization typically required in standard transformer architectures. Experimental results demonstrate that the PT transformer can significantly reduce synchronization overhead while achieving model quality comparable to that of the standard transformer.

\section{The Parallel Track Transformer Architecture}
In this section, we first describe the background on transformers and tensor parallelism for training and inference. We then present the PT transformer architecture. 

\subsection{Background} Transformer-based large language models (LLMs), introduced by~\cite{Vaswani+2017}, have become the backbone of modern natural language processing, leveraging self-attention and feed-forward layers to achieve state-of-the-art performance. These models, often exceeding hundreds of billions of parameters, demand significant computational resources, making single-device execution slow. % "very slow" -> "slow" per review comments

Tensor parallelism~\citep{shoeybi2020megatronlmtrainingmultibillionparameter} addresses this problem by splitting the computation within individual layers across multiple devices. A common form of tensor parallelism is to divide the weight matrices column-wise or row-wise so that each device holds a shard of the full tensor. During computation, each device performs partial operations on its shard, followed by collective communication (all-reduce) to produce the final output. In a standard transformer model, both attention and feedforward layers require synchronization. With an $L$ transformer model, there are a total of $2L$ synchronization points. As $L$ and model size increase, synchronization can become a bottleneck for scaling up LLM inference~\citep{zhang2025laddertransformer}.

\subsection{PT Transformer}
\begin{figure}
    \centering
    \includegraphics[width=0.9\linewidth]{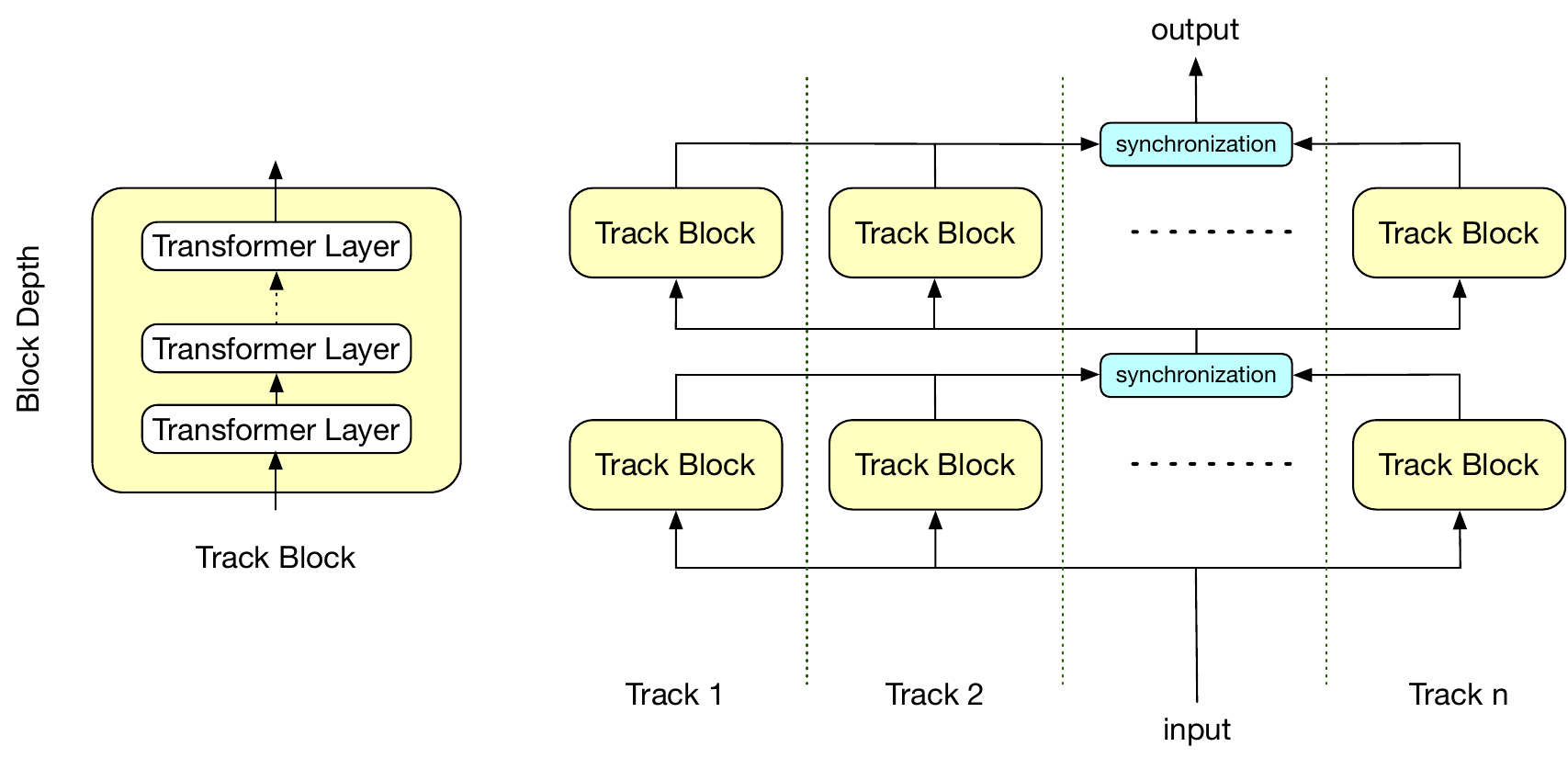}
    \caption{Diagram of the Parallel Track (PT) transformer architecture. Each track is composed of multiple track blocks, and each track block contains a fixed number of standard transformer layers, defined by the block depth.}
    \label{fig:pt-model}
\end{figure}

To reduce the synchronization overhead inherent in tensor parallelism, we propose a new architecture called the PT Transformer. Unlike traditional tensor parallelism—which partitions parameters within individual layers across GPUs to enable parallel computation—the PT Transformer divides the entire model into several smaller transformers, referred to as \textit{tracks} (illustrated in Figure~\ref{fig:pt-model}).

In the ideal scenario where tracks operate completely independently, synchronization is required only at the input and output boundaries of the transformer. However, such isolation typically leads to suboptimal model quality. To address this, we introduce periodic synchronization points between tracks after every \textit{track block}, where each track block consists of a group of standard transformer layers.

Given a total of $L$ transformer layers and a track block depth of $D$, this approach reduces the number of synchronization points from $2L$ to $L/D$. In this design, the depth $D$ effectively controls how often synchronization occurs across GPUs. For example, setting $D = 4$ eliminates $87.5\%$ the synchronization overhead. Moreover, because each individual transformer in PT operates at a reduced dimensionality, the volume of data exchanged during synchronization is also smaller compared to dense models. We refer to this technique as \textit{track parallelism}. The core idea is outlined in Algorithm~\ref{alg:track_paralleism}.

\begin{algorithm}
\caption{Track Parallelism for PT Transformer}
\label{alg:track_paralleism}
\begin{algorithmic}
\State \textbf{Input:} A parallel-track transformer with $n$ tracks. Each track has  $L$ transformer layers with track block depth $D$. Embedding input is $x$.
\State \textbf{Output:} Activation $h$.
\State Set $h_i = x$ for track $i=1,.., n$.
\For{each layer $\ell = 1$ to $L$}
    \State Run $\ell$th transformer layer for each track $i$ in parallel to update $h_i$.
    \If {$\ell \mod  D = 0$} \Comment{Each track block triggers one sync point.}
        \State $h = \textrm{all-reduce}(h_1, ..., h_n) $
    \EndIf
    \State Set $h_i = h$ for track $i=1,.., n$.
 \Comment{Each track receives the same input after all reduce.}
\EndFor
\State Output activation $h$.
\end{algorithmic}
\end{algorithm}

\paragraph{Independent branches with periodic fusion.}
PT is also related in spirit to multi-branch architectures that run several computational ``streams'' in parallel and periodically fuse their representations (e.g., multi-branch transformers with repeated cross-attention fusion~\citep{chen2021crossvit,wang2021crossformer}). A key difference is that, in PT, the fusion operation is not merely a modeling choice but a \emph{systems-motivated} synchronization schedule: tracks are designed to execute largely independently for a fixed block depth $D$, and only then exchange activations via a collective operation. This perspective positions PT as a structured, communication-aware variant of multi-branch transformers where the fusion cadence is explicitly controlled to reduce inter-device dependencies during inference.

\paragraph{Contrast with MoE / expert parallelism.}
Although PT ``tracks'' may superficially resemble experts, PT differs fundamentally from mixture-of-experts (MoE) models~\citep{shazeer2017outrageouslylarge,lepikhin2020gshard,fedus2021switchtransformers}. MoE introduces conditional computation via token-level routing to a (typically sparse) subset of experts, with additional routing decisions and load-balancing considerations; communication is often dominated by dispatch/combine patterns and depends on routing and token distribution~\citep{rajbhandari2022deepspeedmoe,gale2022megablocks}. In contrast, PT does not perform token-level routing: every token is processed by every track, and synchronization occurs at predetermined boundaries (every $D$ layers), yielding a regular communication pattern that is easier to reason about and optimize for serving. Although PT was motivated by inference, introducing track parallelism also offers additional flexibility for improving training efficiency (e.g., enabling new parallelization and scheduling choices that better trade off compute, memory, and communication across devices). In our PT-MoE extension~\citep{zhou2025appleintelligencefoundationlanguage}, MoE sparsity is applied \emph{within} tracks, while track parallelism governs the \emph{cross-device} synchronization schedule.

\section{Experiments}
\label{sec:experiment}
In this section, we begin by outlining our model configurations, followed by an evaluation on several standard benchmarks, and we conclude with a comparison of latency and throughput.

% No throughput comparison version
% In this section, we begin by outlining our model configurations, followed by an evaluation on several standard benchmarks, and conclude with a comparison of time per output token and time to first token.

\subsection{Model Configurations}
We evaluate three model sizes—6B, 13B, and 30B—to compare the performance of our approach against dense baselines. For all PT models, we set the number of tracks to $n = 8$ and adopt Grouped Query Attention (GQA)\citep{ainslie2023gqatraininggeneralizedmultiquery}. See Table~\ref{tab:model-config} for configuration details. In PT models, attention heads and KV heads are evenly distributed across tracks. For instance, in the 30B model, each track is assigned 8 attention heads and 1 KV head, resulting in a total of 64 attention heads and 8 KV heads—identical to the dense model. The 6B model is pretrained on 800B tokens, while the 13B and 30B models are each pretrained on 400B tokens. Both dense and PT models follow the same training recipe.

\begin{table}[h]
\centering
\small
\caption{Model configuration details for dense and PT ($8$ tracks) models.}
\label{tab:model-config}
\begin{tabular}{|c|c|c|c|}
\hline
Model size & \# layers& \# attention heads (per track)&  \# KV heads (per track)\\
\hline
% Example row entries below; replace or add as needed
6B& 32& 32 (4)& 8 (1)\\
13B& 40& 40 (5)& 8 (1)\\ 
30B& 48& 64 (8)& 8 (1)\\
\hline
\end{tabular}
\end{table}

\subsection{Model Performance Comparison}
For PT models, we evaluate all model sizes with track block depths of $D = 2$, $4$, and $8$ against the dense model with the same model sizes. The results for the 6B, 13B, and 30B models are summarized in Table~\ref{tab:model-6b}, Table~\ref{tab:model-13b}, and Table~\ref{tab:model-30b}. As shown in the results, the PT architecture maintains competitive performance compared to standard dense models, even with a track block depth of $D = 8$—which reduces synchronization by 93.75\% (a 16$\times$ reduction). While the 6B model experiences a noticeable drop in MMLU scores as $D$ increases from 2 to 8, larger models such as the 13B and 30B show minimal to no degradation in performance.
\begin{table}[ht]
\centering
\small
\caption{Comparison of model performance for 6B model trained with 800B tokens.}
\label{tab:model-6b}
\begin{tabular}{@{}llcccc@{}}
\toprule
\textbf{Benchmark (Metric)} & \# \textbf{Shots} &
{\textbf{Dense}} &
{\textbf{PT (D=2)}} &
{\textbf{PT (D=4)}} &
{\textbf{PT (D=8)}} \\
\midrule
ARC-C & 0-shot & \textbf{0.492} & 0.472 & 0.476 & \textbf{0.492}  \\
ARC-E & 0-shot & 0.796 & 0.798 & \textbf{0.810} & 0.806  \\
HellaSwag & 0-shot & 0.576 & \textbf{0.580} & 0.578 & 0.574  \\
% LAMBADA & 0-shot & 0.734 & \textbf{0.741} & 0.733 & 0.727  \\
PIQA & 0-shot & 0.791 & 0.794 & \textbf{0.798} & 0.786  \\
SciQ & 0-shot & 0.953 & 0.954 & 0.953 & \textbf{0.955}  \\
WinoGrande & 0-shot & 0.705 & 0.717 & 0.716 & \textbf{0.723}  \\
TriviaQA (EM) & 1-shot & \textbf{0.448} & 0.420 & 0.442 & 0.415  \\
% WebQS (EM) & 1-shot & \textbf{0.259} & 0.190 & 0.247 & 0.240  \\
MMLU (EM) & 5-shot & \textbf{0.560} & 0.548 & 0.514 & 0.360  \\
GSM8K (EM) & 8-shot & 0.317 & \textbf{0.346} & 0.318 & 0.271  \\
MATH (EM) & 4-shot & 0.103 & 0.103 & \textbf{0.113} & 0.094 \\
HumanEval (Pass@1) & 0-shot & 0.139 & 0.171 & 0.161 & \textbf{0.173}  \\
\bottomrule
\end{tabular}
\end{table}

\begin{table}[ht]
\centering
\small
\caption{Comparison of model performance for 13B model trained with 400B tokens.}
\label{tab:model-13b}
\begin{tabular}{@{}llcccc@{}}
\toprule
\textbf{Benchmark (Metric)} & \# \textbf{Shots} &
{\textbf{Dense}} &
{\textbf{PT (D=2)}} &
{\textbf{PT (D=4)}} &
{\textbf{PT (D=8)}} \\
\midrule
ARC-C & 0-shot & 0.476& 0.505& 0.532& \textbf{0.538}\\
ARC-E & 0-shot & 0.795& 0.815& \textbf{0.823}& 0.810\\
HellaSwag & 0-shot & 0.586& 0.590& \textbf{0.591}& 0.584\\
% LAMBADA & 0-shot & \textbf{0.752}& 0.747& 0.751& 0.750\\
PIQA & 0-shot & 0.805& \textbf{0.807}& 0.806& 0.797\\
SciQ & 0-shot & 0.959& 0.961& 0.962& \textbf{0.965}\\
WinoGrande & 0-shot & 0.738& 0.715& 0.729& \textbf{0.743}\\
TriviaQA (EM) & 1-shot & 0.443& 0.441& 0.439& \textbf{0.451}\\
% WebQS (EM) & 1-shot & 0.202& 0.250& 0.242& \textbf{0.253}\\
MMLU (EM) & 5-shot & \textbf{0.583}& 0.576& 0.582& 0.571\\
GSM8K (EM) & 8-shot & 0.374& 0.381& 0.369& \textbf{0.384}\\
MATH (EM) & 4-shot & 0.116& 0.108& 0.111& \textbf{0.118}\\
HumanEval (Pass@1) & 0-shot & \textbf{0.189}& \textbf{0.189}& 0.174& 0.182\\
\bottomrule
\end{tabular}
\end{table}

\begin{table}[ht]
\centering
\small
\caption{Comparison of model performance for 30B model trained with 400B tokens.}
\label{tab:model-30b}
\begin{tabular}{@{}llcccc@{}}
\toprule
\textbf{Benchmark (Metric)} & \# \textbf{Shots} &
{\textbf{Dense}} &
{\textbf{PT (D=2)}} &
{\textbf{PT (D=4)}} &
{\textbf{PT (D=8)}} \\
\midrule
ARC-C & 0-shot & 0.538& 0.536& 0.538& \textbf{0.547}\\
ARC-E & 0-shot & 0.828& 0.837& 0.834& \textbf{0.845}\\
HellaSwag & 0-shot & 0.608& 0.608& 0.609& \textbf{0.610}\\
% LAMBADA & 0-shot & 0.759& 0.769& 0.769& \textbf{0.770}\\
PIQA & 0-shot & 0.809& \textbf{0.811}& 0.808& 0.809\\
SciQ & 0-shot & 0.959& \textbf{0.969}& 0.962& 0.958\\
WinoGrande & 0-shot & 0.735& \textbf{0.768}& 0.750& 0.748\\
TriviaQA (EM) & 1-shot & \textbf{0.487}& 0.478& 0.477& 0.483\\
% WebQS (EM) & 1-shot & 0.261& \textbf{0.291}& 0.274& 0.226\\
MMLU (EM) & 5-shot & 0.630& \textbf{0.646}& 0.602& 0.615\\
GSM8K (EM) & 8-shot & 0.523& \textbf{0.548}& 0.538& 0.488\\
MATH (EM) & 4-shot & 0.168& \textbf{0.174}& \textbf{0.174}& 0.172\\
HumanEval (Pass@1) & 0-shot & 0.223& \textbf{0.262}& \textbf{0.262}& 0.199\\
\bottomrule
\end{tabular}
\end{table}

\subsection{Serving Evaluation}
In this section, we compare a 30B model on 8$\times$H100 across two LLM serving stacks: TensorRT-LLM and vLLM. 
We evaluate PT models with track depths $D\in\{2,4,8\}$ against a dense baseline over diverse input/output lengths.
% Across all settings, by reducing cross-track synchronization, PT improves peak throughput by 15–20\%, reduces both time per output token and time to first token.
% Across all settings, by reducing cross-track synchronization, PT reduced time to first token (TTFT) by 15-30\%, time per output token (TPOT) by 2-12\%, and increased throguhput by up to 31.90\%.
By mitigating cross-track synchronization overhead, across the evaluated settings, PT achieved 15–30\% reduction in time to first token (TTFT), a 2–12\% reduction in time per output token (TPOT), and up to 31.90\% improvement in throughput (with some workload-dependent regressions).

\paragraph{TensorRT-LLM}
% No throughput comparison version
% For TensorRT-LLM, we use an \emph{internal PT-enabled} variant (the open-source release does not support PT at the time of writing). The results are summarized in Table~\ref{tab:TensorRT-LLM-tpot-model-30b}, and Table~\ref{tab:TensorRT-LLM-ttft-model-30b}. As shown, the PT architecture consistently outperforms dense models in time per output token and time to first token in latency mode.

For TensorRT-LLM, we use an \emph{internal PT-enabled} variant (the open-source release does not support PT at the time of writing). The results are summarized in Table~\ref{tab:TensorRT-LLM-throughput-model-30b}, Table~\ref{tab:TensorRT-LLM-ttft-model-30b}, and Table~\ref{tab:TensorRT-LLM-tpot-model-30b}. Bold indicates better performance.  As shown, the PT architecture achieves comparable or superior performance in throughput mode and (for appropriate $D$) improves TTFT and TPOT in latency mode.

\begin{table}[ht]
\centering
\small
\caption{Comparison of TensorRT-LLM model throughput (output tokens per sec) for in throughput mode (maximum batch size = 256)}
\label{tab:TensorRT-LLM-throughput-model-30b}
\begin{tabular}{@{}rrrrrr@{}}
\toprule
\textbf{Input Len} & 
\textbf{Output Len} &
{\textbf{Dense}} &
{\textbf{PT (D=2)}} &
{\textbf{PT (D=4)}} &
{\textbf{PT (D=8)}} \\
\midrule
1024 &  128 & 3193.89 & 3860.32 & 4044.11 & \textbf{4111.98} \\
1024 & 4096 & 4253.77 & 4084.50 & \textbf{4342.59} & 4276.14 \\
2048 &  128 & 1941.80 & 2257.96 & 2415.21 & \textbf{2472.07} \\
2048 & 4096 & 4255.96 & 4003.80 & 4331.89 & \textbf{4350.71} \\
4096 &  128 & 1046.59 & 1281.28 & 1319.82 & \textbf{1344.18} \\
4096 & 4096 & 3154.35 & 3567.53 & 3600.25 & \textbf{3672.80} \\
\bottomrule
\end{tabular}
\end{table}

\begin{table}[ht]
\centering
\small
\caption{Comparison of TensorRT-LLM model TTFT (time to first token) in latency mode (maximum batch size = 1)}
\label{tab:TensorRT-LLM-ttft-model-30b}
\begin{tabular}{@{}rrrrr@{}}
\toprule
\textbf{Input Len} & 
{\textbf{Dense}} &
{\textbf{PT (D=2)}} &
{\textbf{PT (D=4)}} &
{\textbf{PT (D=8)}} \\
\midrule
 1024 &  47.77 &  \textbf{36.04} &  37.35 &  36.64 \\
 2048 &  72.44 &  58.84 &  56.40 &  \textbf{55.30} \\
 4096 &  125.14 &  99.46 &  97.08 &  \textbf{93.75} \\
 8192 &  249.98 &  205.93 &  201.33 &  \textbf{196.61} \\
16384 &  473.69 &  388.71 &  381.10 &  \textbf{377.16} \\
63488 &  1697.20 &  1487.25 &  1452.08 &  \textbf{1436.6} \\
\bottomrule
\end{tabular}
\end{table}

\begin{table}[ht]
\centering
\small
\caption{Comparison of TensorRT-LLM model TPOT (time per output token) in latency mode (maximum batch size = 1)}
\label{tab:TensorRT-LLM-tpot-model-30b}
\begin{tabular}{@{}rrrrrr@{}}
\toprule
\textbf{Input Len} & 
\textbf{Output Len} &
{\textbf{Dense}} &
{\textbf{PT (D=2)}} &
{\textbf{PT (D=4)}} &
{\textbf{PT (D=8)}} \\
\midrule
 1024 &  128 &  6.63 &  6.31 &  6.08 &  \textbf{5.91} \\
 1024 & 4096 &  6.70 &  6.38 &  6.15 &  \textbf{5.98} \\
 2048 &  128 &  6.64 &  6.34 &  6.04 &  \textbf{5.94} \\
 2048 & 4096 &  6.71 &  6.40 &  6.15 &  \textbf{5.99} \\
 4096 &  128 &  6.68 &  6.38 &  6.13 &  \textbf{5.97} \\
 4096 & 4096 &  6.74 &  6.41 &  6.18 &  \textbf{6.01} \\
\midrule
 8192 &  128 &  7.19 &  6.80 &  6.56 &  \textbf{6.44} \\
16384 &  128 &  7.65 &  7.28 &  7.03 &  \textbf{6.91} \\
63488 &  128 &  9.09 &  9.25 &  9.01 &  \textbf{8.88} \\
\bottomrule
\end{tabular}
\end{table}

\paragraph{vLLM}
% No throughput comparison version
% We further evaluated the same set of PT and dense models using vLLM (v0.7.3) in latency mode with the same sequence length configurations. The results, presented in Table~\ref{tab:vllm-tpot-model-30b} and Table~\ref{tab:vllm-ttft-model-30b}, also demonstrate the performance advantage of the PT architecture. % Table~\ref{tab:vllm-throughput-model-30b}

We further evaluated the same set of PT and dense models using vLLM under both throughput and latency modes on the same sequence length configurations.
%The results, presented in Table~\ref{tab:vllm-tpot-model-30b}, also demonstrate the performance advantage of the PT architecture. % Table~\ref{tab:vllm-throughput-model-30b}
We also benchmark 30B size PT models with track block depths of $D = 2$, $4$, and $8$ against the dense model with the same model sizes on different input / output sequence lengths on GPU instances with 8$\times$H100 GPUs. The results are summarized in Table~\ref{tab:vllm-throughput-model-30b}, Table~\ref{tab:vllm-ttft-model-30b}, and Table~\ref{tab:vllm-tpot-model-30b}. Bold indicates better performance.  As shown by the results, the PT architecture yields comparable or better performance in many throughput settings, and consistently improves TTFT and TPOT across the evaluated input/output lengths.

\begin{table}[ht]
\centering
\small
\caption{
Comparison of vLLM model throughput (output tokens per sec) for 30B model in throughput mode (maximum batch size = 256)}
\label{tab:vllm-throughput-model-30b}
\begin{tabular}{@{}rrrrrr@{}}
\toprule
\textbf{Input Len} & 
\textbf{Output Len} &
{\textbf{Dense}} &
{\textbf{PT (D=2)}} &
{\textbf{PT (D=4)}} &
{\textbf{PT (D=8)}} \\
\midrule
1024 &  128 & 2866.34 & 3015.22 & 3079.60  & \textbf{3099.50}  \\
1024 & 4096 & \textbf{5990.98} & 5927.39 & 5839.10 & 5596.01 \\
2048 &  128 & 1632.85 & 1910.94 & 1896.46 & \textbf{1978.98} \\
2048 & 4096 & 5186.72 & \textbf{5224.85} & 5074.91 & 4810.58 \\
4096 &  128 &  865.20 & 1082.73 & 1084.81 & \textbf{1141.18} \\
4096 & 4096 & 4068.51 & \textbf{4278.97} & 4109.63 & 4017.03 \\
\bottomrule
\end{tabular}
\end{table}

\begin{table}[ht]
\centering
\small
\caption{Comparison of vLLM model TTFT (time to first token) for 30B model in latency mode (maximum batch size = 1)}
\label{tab:vllm-ttft-model-30b}
\begin{tabular}{@{}rrrrr@{}}
\toprule
\textbf{Input Len} & 
{\textbf{Dense}} &
{\textbf{PT (D=2)}} &
{\textbf{PT (D=4)}} &
{\textbf{PT (D=8)}} \\
\midrule
 1024 &   69.37 &   59.00  &  58.80  & \textbf{ 54.49} \\
 2048 &  116.98 &  105.37 &  105.42 & \textbf{ 101.75} \\
 4096 &  188.46 &  160.44 &  155.76 & \textbf{ 152.77} \\
 8192 &  337.10 &  288.17 &  271.03 & \textbf{ 261.23} \\
16384 &  646.02 &  530.84 &  510.95 & \textbf{ 501.28} \\
% 32768 & 1361.27 & 1118.96 & 1082.64 & \textbf{1063.03} \\
63488 & 2981.41 & 2543.95 & 2489.36 & \textbf{2452.86} \\
\bottomrule
\end{tabular}
\end{table}

\begin{table}[ht]
\centering
\small
\caption{Comparison of vLLM model TPOT (time per output token) for 30B model in latency mode (maximum batch size = 1)}
\label{tab:vllm-tpot-model-30b}
\begin{tabular}{@{}rrrrrr@{}}
\toprule
\textbf{Input Len} & 
\textbf{Output Len} &
{\textbf{Dense}} &
{\textbf{PT (D=2)}} &
{\textbf{PT (D=4)}} &
{\textbf{PT (D=8)}} \\
\midrule
 1024 &  128 &  8.80 &  8.59 &  8.41 & \textbf{8.42} \\
 1024 & 4096 &  8.88 &  8.67 &  8.52 & \textbf{8.46} \\
 2048 &  128 &  8.87 &  8.63 &  8.51 & \textbf{8.43} \\
 2048 & 4096 &  8.92 &  8.70 &  8.57 & \textbf{8.49} \\
 4096 &  128 &  8.96 &  8.74 &  8.60 & \textbf{8.53} \\
 4096 & 4096 &  9.74 &  9.52 &  9.33 & \textbf{9.24} \\
 \midrule
 8192 &  128 &  9.15 &  8.91 &  8.84 & \textbf{8.76} \\
16384 &  128 &  9.46 &  9.23 &  9.14 & \textbf{9.10} \\
% 32768 &  128 & 10.21 &  9.99 &  9.91 & \textbf{9.85} \\
63488 &  128 & 11.56 & 11.30 & 11.23 & \textbf{11.20} \\
\bottomrule
\end{tabular}
\end{table}

\section{Conclusion}
In this paper, we introduce parallel track transformers, a novel building block within the transformer family that significantly reduces synchronization overhead. We further extend this design by incorporating mixture-of-experts (MoE) to replace the dense MLP layers, yielding the PT-MoE architecture—one of the major advancements highlighted in the Apple Foundation Models 2025 report for private cloud compute~\citep{zhou2025appleintelligencefoundationlanguage}.

%\section*{Author Contributions}
%If you'd like to, you may include  a section for author contributions as is done
%in many journals. This is optional and at the discretion of the authors.

%\section*{Acknowledgments}
%Use unnumbered first level headings for the acknowledgments. All
%acknowledgments, including those to funding agencies, go at the end of the paper.

%\section*{Ethics Statement}
%Authors can add an optional ethics statement to the paper. 
%For papers that touch on ethical issues, this section will be evaluated as part of the review process. The ethics statement should come at the end of the paper. It does not count toward the page limit, but should not be more than 1 page. 

\bibliography{colm2025_conference}
\bibliographystyle{colm2025_conference}

%\appendix
%\section{Appendix}
%You may include other additional sections here.

\end{document}